\documentclass[10pt,preprint]{aastex}

\usepackage{graphicx,epsfig,subfigure,rotating,natbib}
\usepackage{amssymb,amsmath}

\newcommand{\sinc}{{\rm sinc}}

\slugcomment{Accepted to the Astrophysical Journal}

\shorttitle{Effect of Interplanetary Scintillation on EoR Power Spectra}
\shortauthors{C. M. Trott & S.J Tingay}

\begin{document}

\title{The Effect of Interplanetary Scintillation on Epoch of Reionisation Power Spectra}


\author{Cathryn M. Trott\altaffilmark{1,2} \and Steven J. Tingay\altaffilmark{1}}
\affil{International Centre for Radio Astronomy Research, Curtin University, Bentley WA, Australia}

\altaffiltext{1}{ARC Centre of Excellence for All Sky Astrophysics (CAASTRO)}
\altaffiltext{2}{ARC DECRA Fellow}

\begin{abstract}
Interplanetary Scintillation (IPS) induces intensity fluctuations in small angular size astronomical radio sources via the distortive effects of spatially and temporally varying electron density associated with outflows from the Sun. These radio sources are a potential foreground contaminant signal for redshifted HI emission from the Epoch of Reionisation (EoR) because they yield time-dependent flux density variations in bright extragalactic point sources.  Contamination from foreground continuum sources complicates efforts to discriminate the cosmological signal from other sources in the sky. In IPS, at large angles from the Sun applicable to EoR observations, weak scattering induces spatially and temporally correlated fluctuations in the measured flux density of sources in the field, potentially affecting the detectability of the EoR signal by inducing non-static variations in the signal strength. In this work, we explore the impact of interplanetary weak scintillation on EoR power spectrum measurements, accounting for the instrumental spatial and temporal sampling. We use published power spectra of electron density fluctuations and parameters of EoR experiments to derive the IPS power spectrum in the wavenumber phase space of EoR power spectrum measurements. The contrast of IPS power to expected cosmological power is used as a metric to assess the impact of IPS. We show that IPS has a different spectral structure to power from foregrounds alone, but the additional leakage into the EoR observation parameter space is negligible under typical IPS conditions, unless data are used from deep within the foreground contamination region.
\end{abstract}


\section{Introduction}
The solar wind drives the outflow of turbulent plasma into the interplanetary medium, yielding electron density fluctuations that vary spatially and temporally, according to the mean wind velocity. Direct measurements of the density and speed by spacecraft \citep{unti73,neugebauer75,neugebauer76}, and indirect measurements by probing the temporal and spectral fluctuation in the flux density of astronomical radio sources \citep[e.g.,][]{ekers71,bourgois72,armstrong78}, yield a comprehensive understanding of the structure of the medium. Foundation work by \citet{salpeter67,cronyn70,milne76,coles78,readhead78} laid the theoretical groundwork for connecting the power spectrum of density fluctuations to that of phase and intensity fluctuations. 

Interplanetary Scintillation (IPS) can cause the flux density of radio sources in the sky to vary on short timescale. The spatial correlation of sources (observed through the same turbulent interplanetary medium) will cause a systematic increase or decrease in the flux density from a region of the field-of-view. For fluctuation timescales significantly shorter than the observation time (such that the flux density variations average to zero), the only impact will be a short timescale increase or decrease of the system temperature (i.e., the sky temperature). This is a second-order effect compared with direct increase or decrease of spatially-structured power. For intermediary flux density variation timescales (comparable to the observation time), there will be a systematic, spatially-correlated variation in the flux density from a region of sky.

The Epoch of Reionisation (EoR) is a period in the first billion years of the Universe when the first ionising radiation sources heated the intergalactic medium (IGM), and drove the medium phase change from neutral to ionised. Observation of this epoch provides a wealth of information about the spatial and spectral properties of the first ionising sources in the Universe (e.g., Population III stars, black holes etc.). The emission line of neutral hydrogen, at a rest frequency of $\sim$1420~MHz, traces conditions within the IGM by coupling to the gas kinetic temperature and CMB under different ionising conditions \citep{mesinger14,pacucci14,parsons14}. Observation of the redshifted line with low-frequency radio telescopes in the 100--200~MHz band probes $z\sim{6-10}$, and experiments are underway or planned at several telescopes \cite[e.g., Murchison Widefield Array (MWA){\footnote[1]{http://www.mwatelescope.org}}, Precision Array for Probing the Epoch of Reionization (PAPER){\footnote[2]{http://eor.berkeley.edu}}, the Low Frequency Array (LOFAR){\footnote[3]{http://www.lofar.org}}, Long Wavelength Array (LWA){\footnote[4]{http://lwa.unm.edu}}, Hydrogen Epoch of Reionization Array (HERA){\footnote[5]{http://reionization.org}}, Square Kilometre Array (SKA);][]{lonsdale09,tingay13_mwasystem,parsons10,stappers11,ellingson09}. The weakness of the EoR signal relative to the sensitivity of current instruments requires the use of statistical techniques to probe the structure of the signal. The power spectrum, which measures the second moment (variance) of the signal as a function of spatial scale (wavenumber, $k$), is used as the primary detection metric by most of these experiments.

While the spherically-averaged power spectrum (whereby the three spatial dimensions are collapsed into a single wavenumber $k=|\vec{k}|=\sqrt{k_x^2+k_y^2+k_z^2}$) can theoretically yield the greatest detection signal-to-noise ratio by incorporating the most signal, the presence of strong foreground continuum emitters (e.g., radio galaxies, diffuse Galactic emission) contaminate the cosmological signal. Instead, the two-dimensional power spectrum, where the angular and line-of-sight spatial modes remain separate, is used to discriminate cosmological signal from foreground contaminants, using the distinct spectral properties of these two signals. In the former case, spectral structure reflects the line-of-sight distribution in HI brightness temperature. In the latter case, spectral structure maps the spectral energy distribution (SED) of a given source, which is expected to be smooth over small bandwidths ($\sim$~10~MHz). These two signals produce distinct signatures in wavenumber space, allowing discrimination between the signals. The signature of smooth spectrum foregrounds in $k_\bot-k_\parallel$ space is a wedge-like structure, associated with the chromaticity of radio interferometers \citep{datta10,vedantham12,trott12,parsons12,morales12,thyagarajan13,liu14a,liu14b,thyagarajan15,dillon15}. Outside of this wedge is a region named the `EoR Window', which is presumed to be relatively free of foregrounds, and demonstrated as such by recent observational results \citep[e.g.,][]{pober13,dillon15,thyagarajan15,ali15}. This region is used by many experiments in their attempts to measure the cosmological signal (so called, foreground avoidance techniques). Any spectral structure induced by IPS would introduce leakage into the EoR Window. Aside from the overall amplitude of intensity fluctuations, leakage into the window may bias EoR power spectra measurements. For algorithms that operate with a foreground suppression technique, where no data are omitted and foregrounds are treated using a parametric or non-paramteric model, any additional signal due to IPS may be problematic. Recently, IPS was identified in one observation of the EoR with the MWA \citep{kaplan15}.

In Section \ref{sec:approach} we introduce the approach to this problem, before briefly reviewing the key IPS mathematical formalism in Section \ref{sec:formalism} and deriving the foreground contribution to the power spectrum in Section \ref{sec:variance}. A toy model is then used to motivate the derivation in Section \ref{sec:toy}, before the power spectrum of IPS fluctuations is applied to the EoR experiment in Section \ref{sec:ips_eor}. Results are then presented in Section \ref{sec:results}, before discussion and conclusions. Throughout we use a $\Lambda$CDM cosmology with H$_0$=70.4 kms$^{-1}$Mpc$^{-1}$, $\Omega_M$=0.27, $\Omega_k$=0, $\Omega_\Lambda$=0.73 \citep{bennett12}.

\section{Approach}\label{sec:approach}
We aim to compute the bias in measurements of the EoR power spectrum due to fluctuations in sky brightness intensity due to IPS. These intensity fluctuations are caused by turbulent plasma along the line-of-sight to the source distorting the planar wavefront and causing constructive and destructive interference at the observer. In the former case, focussing of light from the background source will increase its apparent flux density, contributing more power from that source than in a static sky model. If the plasma is distributed across the telescope's field-of-view, its structure dictates the effect on the flux density of all sources, yielding a power variation compared with the static case. The structure of the solar wind, and therefore the intensity fluctuations, is characterised by its temporal and spatial power spectrum.

Foreground sources act as a contaminant signal to the underlying cosmological EoR signal, in which we are interested. In general, we can use a two-dimensional power spectrum (separating angular and line-of-sight modes) to disentangle these signals, relying on the assumption that foregrounds are smoothly-varying in frequency, and therefore contained within low line-of-sight wavemodes. This statistical measure of the EoR signal is a primary tool used for signal detection and estimation. The foreground component is a statistical model for the foreground power. Any variation from the static foreground model, due, for example, to IPS-induced intensity fluctuations, will need to be treated in the analysis if the additional power is significant compared with the underlying signal of interest. We address the question of the significance of IPS in this work by (1) using an analytic statistical model for the foreground signal, (2) imprinting the IPS power spectrum upon it, to determine the additional power due to IPS, and (3) comparing the results with the static foreground model and a cosmological EoR model. We compare these results using the spherically-averaged (1D) power spectrum, excising some of the static foreground contribution and quantifying the impact of IPS. We use the 1D power spectrum because it is the metric used by the current generation of EoR experiments, which are attempting to detect the EoR.

We will compare the power due to IPS intensity fluctuations with that due to Galactic and extragalactic emission (in their role as EoR foregrounds), and the expected EoR signal, particularly within the EoR Window. This will serve as the basic input set of observational outputs with which we will assess the relative importance of IPS under normal conditions. More precisely, we will use the relative power of IPS in the spherically-averaged power spectrum compared with a model input cosmological signal, as a metric with which to perform a quantitative comparison.

In the following two sections we develop the formalism for describing a statistical model for extragalactic and Galactic foregrounds (Section \ref{sec:variance}), and motivate the foregoing discussion by providing a simple toy model. We then introduce IPS and motivate its intensity fluctuation power spectrum, based on published results (Section \ref{sec:formalism}), and then imprint the IPS intensity fluctuations on this foreground population (Section \ref{sec:ips_eor}).


\section{Variance of a visibility due to sources in the FOV}\label{sec:variance}
The low-frequency radio sky is bright and dense, with bright emission from extragalactic point sources (e.g., star forming galaxies, AGN), Galactic synchrotron emission, and Galactic compact sources (e.g., supernova remnants). In general, these sources have brightness temperatures orders of magnitude in excess of the cosmological signal \citep[100s~K compared with 10s~mK,][]{jelic08} and present a major challenge for EoR experiments \citep{chapman14,thyagarajan13,thyagarajan15,trott12,bowman09,ali15,patil14}. For the purposes of this study, we need to understand the effect of IPS on these foregrounds, because these are the sources of contamination to our signal. Here, we develop the basic formalism for building a statistical model of extragalactic point sources and Galactic synchrotron emission, and then imprint the intensity fluctuations due to IPS upon these. Note that because Galactic synchrotron emission has most power on large spatial scales, we expect that IPS will have less impact than on the unresolved sources, however there is still the possibility of spectral structure due to IPS.

\subsubsection*{Extragalactic point sources}
The power spectrum metric computes the power (variance) on each spatial scale, integrated across the sky. The foreground contribution to the power spectrum due to point sources is due to the imbalance of the distribution of sources across the sky, whereby the number of sources of a given flux density in a unit area of sky is Poisson-distributed ($\mathcal{P}()$):
\begin{equation}
N(S,S+dS) dS\,dl\,dm \sim \mathcal{P}(\langle{N}\rangle),
\end{equation}
where $\langle{N}\rangle$ is the expected number, given parametrically by:
\begin{equation}
\langle{N(S,S+dS)}\rangle(\nu) = \frac{dN}{dS}(\nu)\,dS\,dl\,dm = \alpha \left( \frac{\nu}{\nu_0} \right)^{\gamma} \left( \frac{S_{\rm Jy}}{S_0}\right)^{-\beta}\,dS\,dl\,dm.
\label{source_counts}
\end{equation}
We use values of $\alpha=4100\,{\rm Jy}^{-1}{\rm sr}^{-1}$, $\beta=1.59$ and $\gamma=0$, in line with recent measurements \citep{intema11,gervasi08}.

If sources were distributed evenly across the sky, all interferometric visibilities would be identically zero (in the absence of noise and with a symmetric, infinite primary beam). However, the truncation and attenuation of the sky by the telescope beam (field-of-view) allows power to be leaked from the autocorrelation mode ($k=0$) into non-zero wavenumbers (the short spacing problem). In addition to this power from the \textit{expected} number of sources in the sky, the number at any given location in the sky is Poisson-distributed, leading to variance in the measured value of a visibility due to the random imbalance between sources. This foreground covariance term corresponds to one component of the wedge of contamination observed in the EoR power spectrum. For a given visibility, the covariance between spectral channels can be derived from the distribution of signal in visibilities, and the expectation that differential regions of the sky are mutually independent, such that:
\begin{equation}
{\rm Var}[N(<S_{\rm max})]\,dl\,dm = \int^{S_{\rm max}}_0 dS\,S^2 \frac{dN}{dS} \,dl\,dm, 
\end{equation}
gives the variance of the number of sources with flux density $S<S_{\rm max}$ per differential sky area. The total covariance is then:
\begin{eqnarray}
\boldsymbol{C}_{\rm FG} &\equiv& \langle (\vec{V} - \langle\vec{V}\rangle)^\dagger (\vec{V} - \langle\vec{V}\rangle)  \rangle\\
&=& \iiint S^2 \left( \frac{\nu}{\nu^\prime} \right)^{-\gamma} B(l,m;\nu) B(l,m;\nu^\prime) \frac{dN}{dS}dSdldm \\
&=&  \frac{\alpha}{3-\beta} \left( \frac{\nu}{\nu^\prime} \right)^{-\gamma} \frac{S_{\rm max}^{3-\beta}}{S_0^{-\beta}} \iint B(\vec{l};\nu)B(\vec{l};\nu^\prime) \exp{-2\pi{i}(\vec{u}\cdot\vec{l})(\nu-\nu^\prime)} d\vec{l} \hspace{0.3cm} {\rm Jy^2},
\end{eqnarray}
where $S_{\rm max}$ is the brightest unmodelled source in the field (the peeling limit), $B(l)$ is the beam response, and $\alpha$ and $\beta$ parametrise the source number counts as in Equation \ref{source_counts}.
For a single channel, the variance reduces to:
\begin{equation}
\boldsymbol{C}_{\rm FG} = \frac{\alpha}{3-\beta} \frac{S_{\rm max}^{3-\beta}}{S_0^{-\beta}} \iint B^2(\vec{l};\nu) d\vec{l}.
\end{equation}
For a beam that can be approximated by a Gaussian ($\propto \exp{(-l^2/\sigma^2)}$), the covariance matrix between spectral channels can be computed analytically, and is:
\begin{equation}
\boldsymbol{C}_{\rm FG}(\nu,\nu^\prime;k) = \frac{\alpha}{3-\beta} \Gamma(\nu)\Gamma(\nu^\prime) \left( \frac{\nu}{\nu^\prime} \right)^{-\gamma} \frac{S_{\rm max}^{3-\beta}}{S_0^{-\beta}} \frac{\pi{c^2}\epsilon^2}{D^2}\frac{1}{\nu^2 + \nu^{\prime{2}}} \exp{\left( \frac{-k^2c^2f(\nu)^2\epsilon^2}{4(\nu^2 + \nu^{\prime{2}})D^2} \right)},
\end{equation}
where $\sigma \simeq \frac{\epsilon{c}}{\nu{D}}$ and $\nu=$~180~MHz is the observation frequency, $\epsilon=0.42$, $D=4$~m are the scalings from an Airy disk to a Gaussian width, and the MWA tile diameter, respectively, giving $\sigma\simeq{0.17}$ radians, and $f(\nu)\equiv (\nu_2-\nu_1)/\nu_0$. The $\Gamma(\nu)$ functions contain any bandpass shaping (window taper function) applied prior to Fourier Transform to reduce spectral leakage. In this work we use a Blackman-Nuttall window, to reduce sidelobe contamination \citep{nuttall81}. The impact of this choice is explored below. For a given frequency channel, the variance between spatial wavenumber $k$ and frequency $\nu$ is:
\begin{eqnarray}
\boldsymbol{C}_{\rm FG}(\nu;k) &=& \frac{\alpha}{3-\beta} \frac{S_{\rm max}^{3-\beta}}{S_0^{-\beta}} \frac{\pi\sigma^2}{2}\\
&:=& g(S) \frac{\pi\sigma^2}{2} \,\,{\rm Jy^2}. 
\label{variance}
\end{eqnarray}
Here we have defined the function $g(S)$.
Equation \ref{variance} gives the variance of a visibility due to well-behaved sources distributed across the sky (i.e., non-IPS-affected sources).

\subsubsection*{Galactic synchrotron}
We follow the simple parametric model from \citet{jelic08} to represent the power spectrum of Galactic synchrotron emission. This model follows a power law in frequency and wavenumber, with a steep index in the latter. The intrinsic temperature power spectrum is modelled as:
\begin{equation}
\langle \Delta{T}^2_{\rm GS}\rangle(k,\nu) = (\eta{T_B})^2\,\left(\frac{k}{k_0}\right)^{-2.7}\left(\frac{\nu}{\nu_0}\right)^{-2.55}\,\,{\rm K^2},
\end{equation}
where $\eta=0.01$ is the fluctuation level relative to the uniform brightness temperature, $T_B=253{\rm K}$, $k_0 = 10$ wavelengths, and $\nu_0=100$~MHz is the reference frequency. The apparent steep spectral index in temperature units is flattened once converting to flux density (integrated) units for a given instrument, such that the intrinsic power is:
\begin{eqnarray}
\boldsymbol{C}_{\rm GS}(\nu,\nu^\prime;k) &=& \left(\frac{2k}{\lambda\lambda^\prime}\right)^2\Omega(\eta{T_B})^2\,\left(\frac{k}{k_0}\right)^{-2.7}\left(\frac{\sqrt{\nu\nu^\prime}}{\nu_0}\right)^{-2.55}\,\,{\rm Jy^2}\\
&=& \left(\frac{(2k)^2}{\lambda\lambda^\prime}\right)\frac{1}{A_{\rm eff}}(\eta{T_B})^2\,\left(\frac{k}{k_0}\right)^{-2.7}\left(\frac{\sqrt{\nu\nu^\prime}}{\nu_0}\right)^{-2.55}\,\,{\rm Jy^2}\\
&=& \left(\frac{(2k)^2}{A_{\rm eff}}\right)(\eta{T_B})^2\,\left(\frac{k}{k_0}\right)^{-2.7}\left(\frac{\sqrt{\nu\nu^\prime}}{\nu_0}\right)^{-0.55}\,\,{\rm Jy^2}.
\end{eqnarray}

The full foreground covariance is given by the sum of the two components.

\subsubsection*{Experimental design}
For this work, we model an instrument matching the general design of the MWA, but with a matched-size frequency-dependent Gaussian beam shape replacing the actual beam, for simplicity of applying the analytic model. The beam has been described above. The sensitivity and array layout of the instrument is not considered here, and therefore the results may be applied generically to any low-frequency widefield telescope. The parameters of the MWA EoR experiment \citep{jacobs15} are used as input: $z=8.2$, $\nu_{\rm base}=150{\rm MHz}$, BW$=30.72{\rm MHz}$, $N_{\rm chan}=384$, giving a spectral channel resolution of $\Delta\nu=40{\rm kHz}$. Sources are assumed to have been subtracted perfectly to a flux density limit of $S_{\rm max}=100$~mJy.

The Blackman-Nuttall window tapering function has been applied in this work (represented as $\Gamma(\nu)$ above). This is because it offers a high degree (10-dex in power) of sidelobe suppression, which is useful for `foreground avoidance' techniques where the foreground contamination region is excised from the analysis. For more general techniques, the tapering function does not serve to improve the detectability of the signal, because the same taper is applied to static foregrounds, IPS power and the cosmological signal. One disadvantage of using the Blackman-Nuttall is its broad main lobe, which occupies four of the lowest wavemodes (compared with no taper, which occupies one, and the Hanning Window, which occupies two). This pushes the contaminated wedge region further into the parameter space, but with the benefit of less leakage into higher modes.

\section{Toy model}\label{sec:toy}
To illuminate the problem, we construct a simplified toy model, whereby a single source at the phase centre experiences an IPS-induced intensity fluctuation, similar in style to the observation of IPS by the MWA \citep{kaplan15}. We demonstrate the size of this effect, relative to the foreground covariance power from all extragalactic point sources. The source has a flux density set to equal the limit below which sources are not individually calibrated, because sources brighter than this will be measured and any IPS fluctuations identified. For the MWA, this limit is currently set at $S_{\rm int}$~=~300~mJy at zenith.

We consider a single source, that exhibits an IPS-induced flux density fluctuation of fractional amplitude $\Delta{F}$, such that:
\begin{equation}
S_{\rm obs} = (1 + \Delta{F})S_{\rm int},
\end{equation}
is the observed flux density. For a single source, the expected value of a visibility due to enhancement by IPS is:
\begin{equation}
\langle V(u,v) \rangle = S_{\rm int} \Delta{F} B(l,m;\nu) \exp{-2\pi{i}(\vec{u}\cdot\vec{l})}.
\end{equation}
We will take the most pessimistic approach and assume that the IPS is affecting the region of the beam with the highest sensitivity (close to the boresight, $B\simeq{1}$), giving:
\begin{equation}
\langle V(u,v) \rangle \simeq S_{\rm int} \Delta{F} \,\,{\rm Jy}.
\label{expsq}
\end{equation}
To compare with the variance of a visibility, we take the square of the expected value. Equating \ref{variance} and \ref{expsq} gives:
\begin{equation}
\frac{\alpha}{2-\beta}\frac{S_{\rm int}^{2-\beta}}{S_0^{-\beta}}\frac{\pi\sigma^2}{2} = S_{\rm int}^2(1+\Delta{F})^2
\end{equation}
yielding,
\begin{eqnarray}
\Delta{F}_{\rm min} &=& \sqrt{\frac{h(S)\pi\sigma^2}{2}}\frac{1}{S_{\rm int}}-1\\
&\approx& 50,
\end{eqnarray}
as an indicative minimum fractional fluctuation in source flux density for the power to exceed that from static extragalactic point sources (where $h(S) \equiv \frac{\alpha}{2-\beta}\frac{S_{\rm int}^{2-\beta}}{S_0^{-\beta}}$).

This minimum value exceeds, by a factor of a few, the observation of IPS described in \citet{kaplan15}, which showed variation in two sources by factors of 10--20 over 2~minute intervals. No other sources in the field showed statistically significant variability. Note that these sources were bright, with flux densities of several Jansky. These sources were identified because they are bright and did not match with their catalogue value. For EoR observations, these sources would be calibrated and excised from the data, and it is only weaker sources that would propagate power through to the calibrated dataset. \citet{trott12} demonstrated that residuals from fitting and subtracting of bright point sources were small compared with power from unsubtracted sources, and any additional IPS-induced intensity fluctuations in these would therefore be at a level further below this. It is not, however, observed that all sources exhibit such large fractional flux density variations. \citet{loi15} studied EoR fields from the MWA experiment for scintillation, and found typical scintillation indices, $m=1-3\%$, using 51 hours of data due to all sources of scintillation (ionospheric, IPS). These factors suggest that modulation indices exceeding 10s are rare. We now proceed to develop a more rigorous framework for the temporal and spatial power spectrum of IPS, under normal conditions, and apply that to the entire field of foreground sources.


\section{Interplanetary Scintillation formalism}\label{sec:formalism}
In this section we rely on existing literature and measurements of IPS to build a power spectrum of IPS-induced intensity fluctuations, which we can imprint onto our foreground statistical model, developed in Section \ref{sec:variance}.

The theoretical basis for scintillation studies was developed during the middle of the 20th Century, with a large body of research produced in the 1960s and 1970s. Here we reproduce the key results to motivate the present work, and refer to relevant publications.

IPS is the fluctuation in wavefront phase (and hence intensity) received by the observer due to interaction of a planar wave with plasma density variations in the interplanetary medium. In the regime of weak scattering, ray bundles are deflected and compressed by fluctuations in the electron density, yielding phase delays along different lines-of-sight. In the interplanetary medium, electron density fluctuations are caused by the radial outflow of the turbulent solar wind \citep{salpeter67,cronyn70,coles78}. At the observer plane, these distortions yield weak scintillation of the flux density of sources. At large solar elongations (angular separation of the observation field from the Sun) typical of Epoch of Reionisation experiments where wavefront phase distortions are $\lesssim$~1 rad, weak scintillation is the dominant source of intensity fluctuations, whereby the signal wavefront is distorted by weak scattering.

To derive the power spectrum of intensity fluctuations, we follow the literature and begin with spacecraft measurements of the density power spectrum, and link this to phase fluctuations and intensity fluctuations. For EoR datasets, which are measured at large elongation to the Sun, we are in the weak scattering regime, and use approximations to derive the intensity power spectrum. In the following section we will describe the additional impact on the power spectrum of the temporal and spatial sampling of the telescope.

Following \citet{salpeter67} and \citet{bourgois72}, the three-dimensional electron density power spectrum is given by the Fourier Transform of the spatial two-point correlation function:
\begin{equation}
F_{ns}(k_x,k_y,k_z) = \iiint d^3{x} \exp{(-i\vec{k}\cdot\vec{x})} \langle n(\vec{x})n(\vec{x}+\vec{r}) \rangle,
\end{equation}
where $n$ is the electron density, in number per volume, and the subscript denotes $n$=electron density, $s$=spatial power. We define a Cartesian co-ordinate system where $x,y$ lie transverse to the line-of-sight, $z$. 

Due to the directional nature of the flow of plasma from the Sun, the electron density spectrum can be anisotropic, reflecting anisotropy in the underlying turbulence, and leading to potential large deviations from the scintillation properties of an isotropic distribution \citep{macquart07}. In this case, an axial ratio, $R$, of electron density may be introduced, which characterises elongation of the turbulent structures. The power spectrum of IPS is observed to be anisotropic \citep{dennison69,cronyn70,malaspina10}. We extend the isotropic electron density power spectrum to include an elongation (axial ratio) parameter, $R>1$, and angle, $\theta$, over elongation to the $x$ axis (direction of motion of solar wind) \citep{rickett02,macquart07}. In line with observations, we take $\theta=0$, and define the anisotropic power spectrum \citep{cronyn70,milne76,coles78}:
\begin{equation}
F_{ns}(k_x,k_y,k_z) = C_N^2\Delta{L}\,\left( k_x^2/R + Rk_y^2 + k_z^2 \right)^{-n/2}\,\,{\rm m}^{-3} ,
\label{density_ps_model}
\end{equation}
where $n\simeq{3.5}$. We introduce the concept of a phase screen, whereby the wavefront distortions are approximated to occur over a fixed distance. In the thin screen approximation, if the amplitude, $C_N^2(z)$, evolves slowly with $z$, then the line-of-sight integral over the electron density power spectrum is approximately,
\begin{equation}
{\rm SM} \equiv C_N^2\Delta{L} \,\,{\rm m}^{-2},
\end{equation}
where SM denotes the scattering measure, and $\Delta{L}$ is the screen thickness.

The electron density fluctuations yield refractive index fluctuations, leading to phase fluctuations in the incoming wavefront:
\begin{equation}
F_{\phi{s}}(k_x,k_y) \simeq 2\pi({r_e}\lambda)^2\Delta{L} F_{ns}(k_x,k_y,0),
\end{equation}
where taking the $k_z=0$ mode yields a flat integration along the line-of-sight over density fluctuations, $r_e$ is the classical electron radius, and $\lambda$ is the observing frequency.

In the weak scattering regime, the intensity fluctuation spatial power spectrum is the product of the phase power spectrum and a high-pass filter (the Fresnel filter), which suppresses large scale (small $k$) power:
\begin{equation}
F_{Is}(k_x,k_y) = F_{\phi{s}}(k_x,k_y).F_F(k_x,k_y),
\end{equation}
where
\begin{eqnarray}
F_F(k_x,k_y) = 4\sin^2{k^2/k_F^2},\\
k_F^2 = \frac{4\pi}{\lambda{L}}\,\,{\rm m}^{-2}
\end{eqnarray}
attenuates scales where refraction is insufficiently strong for interference to occur, and $L$ is the distance to the phase screen. The filter defines the spatial scale corresponding to the first Fresnel zone, and defines the region larger than which the wavefront distortions are not sufficiently strong to produce interference. The Fresnel frequency at 150~MHz is $k_F\simeq{10}^{-4}$~km$^{-2}$.

The solar wind moves at speed $U$ radially from the Sun, yielding a temporal power spectrum in one dimension that can be derived from the spatial power spectrum. If we define the direction of motion to be $x$, the temporal power spectrum is obtained by integrating over all $y$ spatial modes:
\begin{eqnarray}
F_{It}(\omega) &=& 2\pi(\lambda{r}_e)^2\,\Delta{L}\, \int\, dk_y \, F_{Is}(k_x,k_y)\\
&=& 2\pi(\lambda{r}_e)^2 \frac{\Delta{L}}{U}\, \int\, dk_y \, F_{Is}(\omega/U,k_y)
\label{f_it}
\end{eqnarray}
with $\omega = k_x{U}$ giving the temporal angular frequency associated with an $x$-direction spatial scale. The temporal power spectrum is required for normalising the electron density power spectrum to measurements of the scintillation index of sources, $m$. The scintillation index measures the rms flux density of a source relative to its mean value, and corresponds to the integral over the temporal power spectrum (i.e., the zero-lag term of the temporal autocorrelation function; this corresponds to the fractional flux density, $\Delta{F}$, used in the toy model):
\begin{equation}
m^2 = \int d\omega\,F_{It}(\omega).
\label{scintillation_index}
\end{equation}
\citet{readhead78} provide measurements of the scintillation index of extragalactic sources as a function of solar elongation. For angles $\geq$~20 degrees, $m^2\leq{0.4}$. They fit an empirical relationship to data, which gives the scintillation index as a function of observing frequency, $\nu$, and solar elongation distance, $\xi$:
\begin{equation}
m = 23\,\left(\frac{{\rm 150~MHz}}{\nu}\right) \left(\frac{\xi}{1.{\rm AU}}\right)^{-1.5},
\label{m_elon}
\end{equation}
where $1\, {\rm AU}=1.5\times{10}^{11}{\rm m}$.

The scintillation index is used to fix the value for the scattering measure, $C_N^2\Delta{L}$. Connecting Equations \ref{f_it} and \ref{scintillation_index} yields:
\begin{equation}
m^2 =  2\pi(\lambda{r}_e)^2 \frac{4\Delta{L}C_N^2}{U}\,\int d\omega \, \int dk_y\, k^{-n} \,\sin^2{(k^2/k_F^2)},
\label{intensity_ps}
\end{equation}
where $k^2 = Rk_y^2 + ({\omega}/U)^2/R$.


\section{Observed IPS power spectrum}\label{sec:ips_eor}
We use Equation \ref{m_elon} in conjunction with Equation \ref{intensity_ps} to compute the integrated amplitude of the electron density fluctuations, $C_N^2\Delta{L}$. Once set, we have a full description of the spatial power spectrum of IPS intensity fluctuations as a function of observing frequency, for a \textit{perfect} instrument (i.e., for adequate temporal sampling of the fluctuations to avoid aliasing). \textit{Instead}, we have an instrument that averages data over a given time interval, $\Delta{T}$, given by the visibility averaging timescale. This timescale acts as a low-pass filter, suppressing temporal variability (and therefore spatial variability in the direction of motion of the solar wind) on shorter timescales (larger $k$-modes). For a rectangular temporal window, the Fourier response is a sinc\footnote{$sinc(x) = \frac{\sin{(\pi{x})}}{\pi{x}}$} function, and we vary the intensity power spectrum, Equation \ref{intensity_ps}, such that:
\begin{equation}
P_{\rm obs}(k_x,k_y;\nu) = 8\pi\Delta{L} \left(\frac{cr_e}{\nu}\right)^2 C_N^2 \left( \frac{\sqrt{k_x^2/R+Rk_y^2}}{k_0} \right)^{-n} \sin^2{\left(\frac{k_x^2+k_y^2}{k_F(\nu)^2}\right)} \sinc^2{\left(\frac{Uk_x\Delta{T}}{2}\right)}\,{\rm sr}.
\label{final_intensity_ps}
\end{equation}
Here, the $x$ and $y$ modes have slightly different dependencies due to the sinc smearing affecting only the direction of plasma motion and the anisotropy of the power spectrum.

Finally, we combine Equation \ref{final_intensity_ps} with the source flux density distribution function, $h(S)$ (using Equation \ref{expsq}), to give the IPS variance:
\begin{equation}
\boldsymbol{C}_{\rm IPS}(\nu,k) = h(S)^2 P_{\rm obs} \frac{\pi\sigma^2}{2} \,{\rm Jy}^2.
\end{equation}
Again, this can be compared with the foreground variance, Equation \ref{variance}, and a simple model for a cosmological power spectrum of 21~cm brightness temperature fluctuations. Compared with the toy model, the inclusion of the correct IPS power spectrum yields a scale-dependent  characteristic `area', where before we considered an isolated source.

For this experiment, we assume that the density fluctuations are at a distance of 1~AU, travelling at 350~km~s$^{-1}$ \citep{cronyn70,readhead78} and set $m=1$.

\subsection{Connection to EoR $k$-space}
Having derived the variance (power) of the IPS in $k-\nu$ space, the final steps are to, (1) Fourier transform the frequency axis to obtain the line-of-sight modes, and (2) convert the solar system-defined units to the cosmological scales used by the EoR. In step (1), we operate on the signal covariance to obtain the $k-\eta$ variance, taking full account of the spectral correlations in the frequency covariance matrix:
\begin{equation}
\boldsymbol{C}(\eta,\eta^\prime;k) = \left(\mathcal{F}^\dagger \boldsymbol{C}(\nu,\nu^\prime;k) \mathcal{F}\right)  \,(\Delta\nu)^2 \,\,{\rm Jy}^2{\rm Hz}^2,
\label{ft}
\end{equation}
where $\mathcal{F}$ is the DFT matrix and $\Delta\nu$ is the spectral resolution of a given channel to keep consistency of units.

In step (2), we use the following scalings from SI to cosmological units \citep{morales04}:
\begin{eqnarray}
k_\bot &=& \frac{2\pi{|\boldsymbol{u}|}}{D_M(z)}\,{\rm Mpc}^{-1},\\
k_\parallel &=& \frac{2\pi{H_0}f_{21}E(z)}{c(1+z)^2}\eta\,{\rm Mpc}^{-1},\\
1 {\rm Jy^2.Hz}^2 &=& 10^{-20}\, \frac{2kT}{\Omega\lambda^2} {\rm V}_{\rm Mpc^3\,sr^{-1}\,Hz^{-1}}\,\, {\rm mK^2\,Mpc^3},
\end{eqnarray}
where $D_M(z), H_0, f_{21}, z$ are the transverse comoving distance, Hubble constant, rest frequency of the hyperfine transition ($\sim$1420~MHz), and observation redshift, respectively, and $V$ is the conversion from SI to cosmological units \citep{mcquinn06}. $E(z)$ is \citep{hogg99}:
\begin{equation}
E(z) \equiv \sqrt{\Omega_M(1+z)^2 + \Omega_k(1+z)^2 + \Omega_\Lambda}.
\end{equation}
Beyond the 2D power spectrum, we average the power in cylindrical shells and normalise by the volume sampled by each spatial mode to obtain the spherically-averaged dimensionless power spectrum:
\begin{equation}
\Delta^2(k) = \frac{k^3}{2\pi^2}P(k)\,\,{\rm mK^2},
\label{equation:1d}
\end{equation}
where, in its simplest form with equal weights, $N_k$ bins contributing to wavenumber $k$, and observation volume, $\Omega$, the 1D power is,
\begin{equation}
P(k) = \frac{1}{N_k}\displaystyle\sum_{i \in k} P_i(k_\bot,k_\parallel) = \frac{1}{\Omega{N_k}}\displaystyle\sum_{i \in k} \boldsymbol{C}(\eta,\eta;k)_i(k_\bot,k_\parallel)\,\,{\rm mK^2\,Mpc^3}.
\end{equation}

\section{Results}\label{sec:results}
\begin{figure}
\plotone{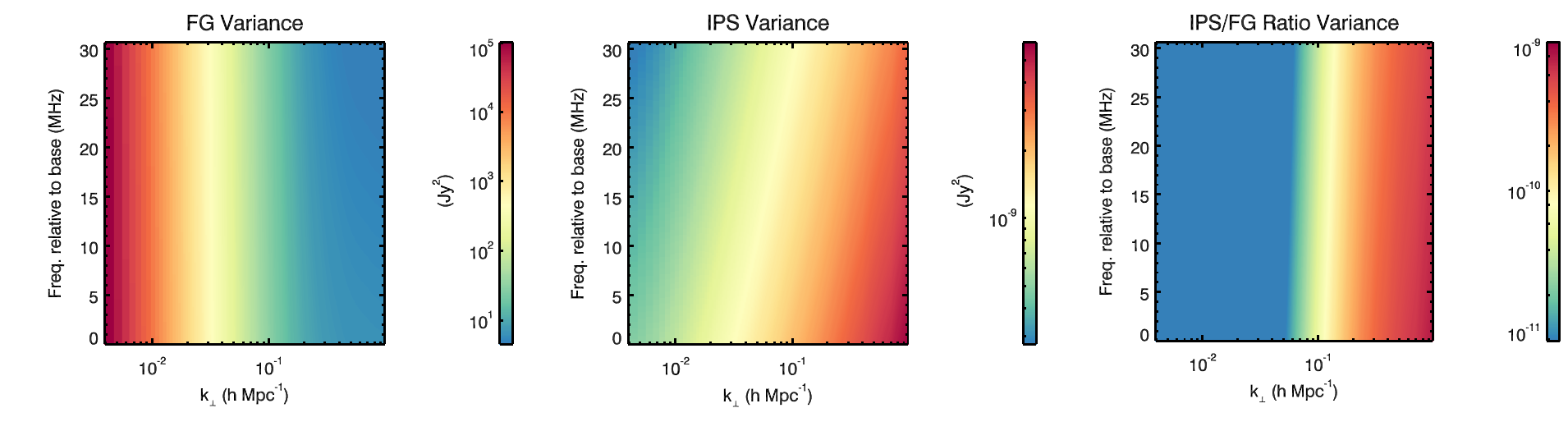}
\caption{ (Left) Variance of foregrounds as a function of angular scale and frequency; (centre) IPS variance; (right) ratio of IPS:foreground contribution to the power.\label{fig:variance}}
\end{figure}

Figure \ref{fig:variance} shows the $k-\nu$ space variance of the foregrounds (left panel) and IPS (centre panel), respectively, in units of ${\rm Jy}^2$ for the isotropic case ($R=1$). The right panel displays the ratio. No bandpass taper has been applied at this stage.
The frequency dependence of the Fresnel filter introduces additional spectral structure into the IPS power, compared with the foregrounds, and the strong wavenumber-dependence of the Galactic synchrotron emission is evident. Note that IPS power is concentrated on small spatial scales, yielding a much stronger impact on point sources than Galactic synchrotron emission.

\begin{figure}
\plotone{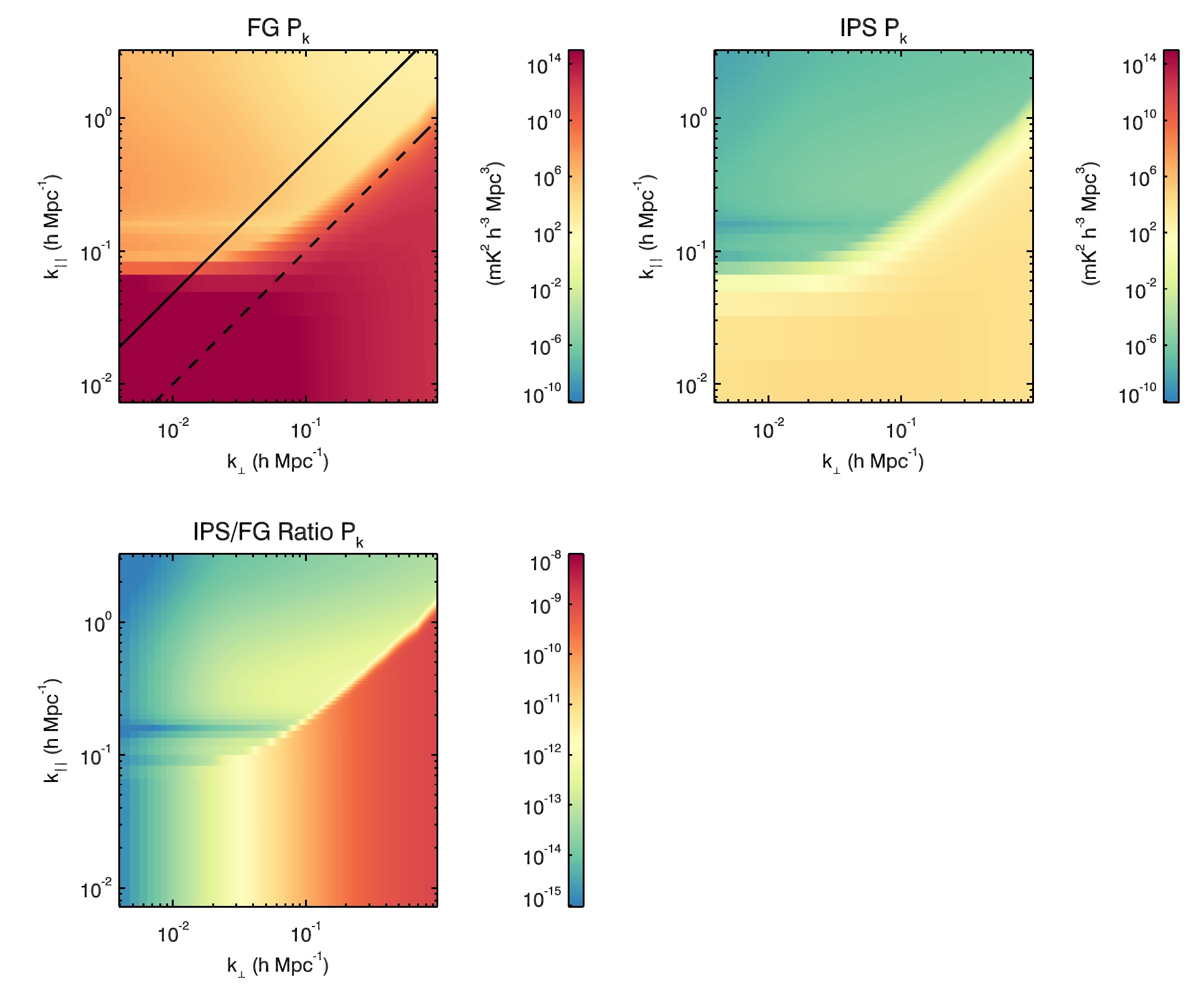}
\caption{ (Top left) Power spectrum of foregrounds in EoR wavenumber phase space, $k_\bot-k_\parallel$; (top right) IPS power; (bottom left) ratio of IPS:foreground contribution to the power. The horizontal bands at $k_\parallel \approx 0.15{\rm h~Mpc}^{-1}$ are due to nulls in the Blackman-Nuttall taper function.\label{fig:power}}
\end{figure}

Figure \ref{fig:power} uses Equation \ref{ft} to take the frequency-space covariance matrices and transforms them to line-of-sight wavenumber space, using a range of angular $k$ modes ($k_\bot$) consistent with a typical EoR experiment, in units of ${\rm mK}^2\,h^{-3}{\rm Mpc}^3$.
The wedge of emission expected for foreground sources is visible in low $k_\parallel$ modes, and the Galactic synchrotron component is visible at low $k_\bot$ modes. The leakage of power beyond the low $k_\parallel$ modes expected for smooth-spectrum foreground components is due to the finite bandwidth of the experiment (spectral leakage), and dependent on the choice of bandpass tapering function to suppress this. These features are also present for the IPS power, but there additional spectral structure leaks power further into the EoR window. Although the ratio of IPS to foreground power in the EoR window is low, there is excess power within the foreground dominated region that may require additional treatment by EoR experiments. The horizontal bands at $k_\parallel \approx 0.15{\rm h~Mpc}^{-1}$ are due to nulls in the Blackman-Nuttall taper function. In the ratio, these bands occupy slightly different wavemodes due to the interaction between them and the differing spectral structure of the static and IPS foregrounds. A different choice of taper function would alter these plots somewhat, but the general conclusions remain unchanged.

\begin{figure}
\plotone{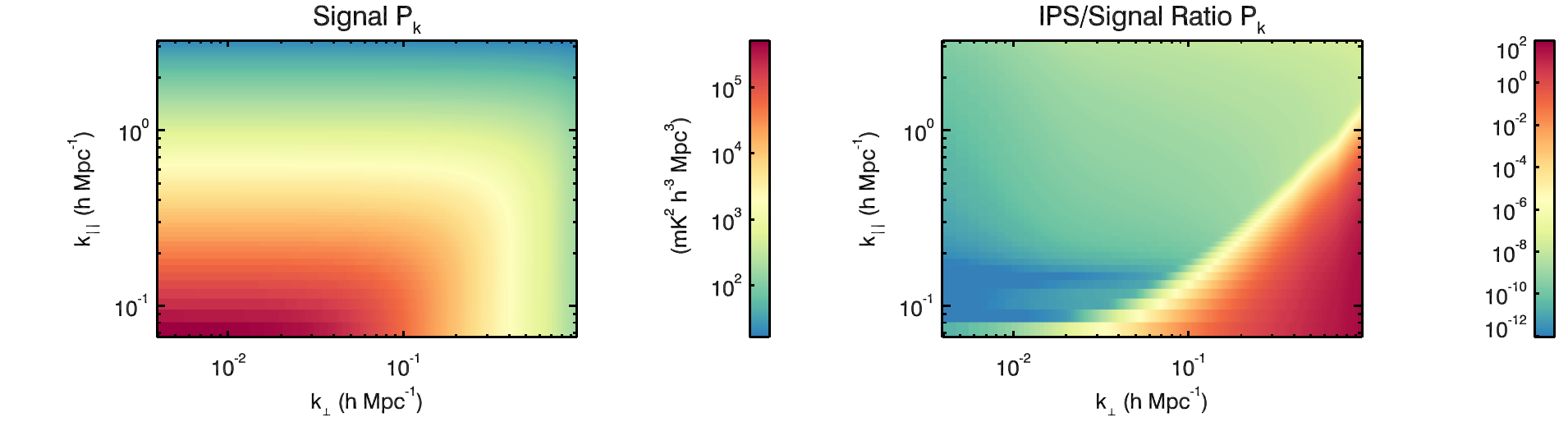}
\caption{ (Left) Expected power spectrum of cosmological brightness temperature fluctuations ($z=8$, $x_{\rm HI}=1$); (right) ratio of IPS to cosmological signal.\label{fig:signalpower}}
\end{figure}

We now compare the IPS power amplitude to that expected for a typical model of the cosmological power spectrum at $z=8$ and for a fully-neutral medium \citep{furlanetto06,eisenstein99}. Figure \ref{fig:signalpower} shows the expected signal power (left panel) and the ratio of IPS to 21~cm signal power (right panel).
Within the wedge, there is contamination that exceeds the expected 21~cm signal power. Outside of the wedge, the effects of IPS are estimated to be orders of magnitude smaller than the signal of interest.

\subsubsection*{Impact on the spherically-averaged power spectrum}
\begin{figure}[t]
\centering
\subfigure{
\includegraphics[width=.36\textwidth]{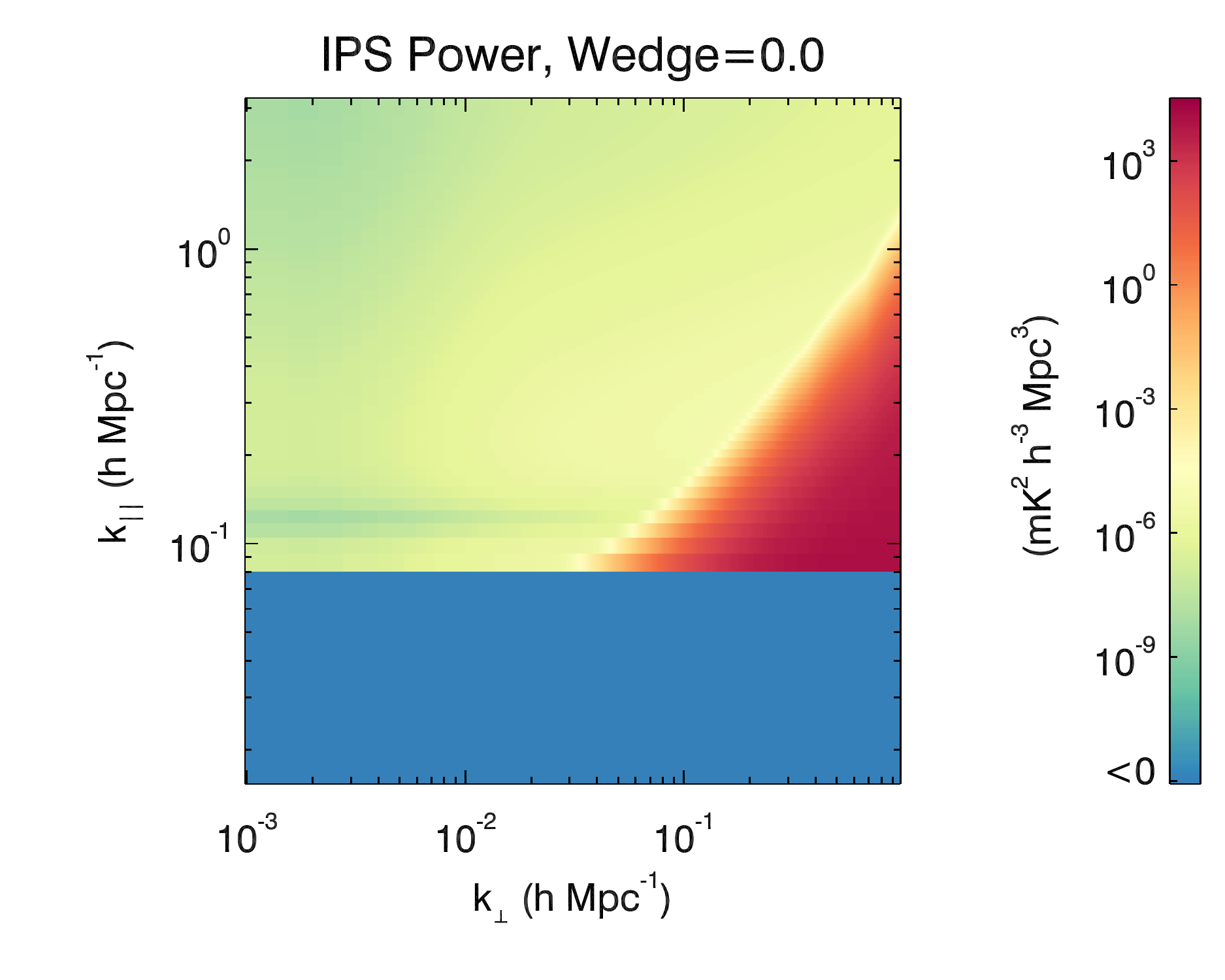}
}
\subfigure{
\includegraphics[width=.28\textwidth]{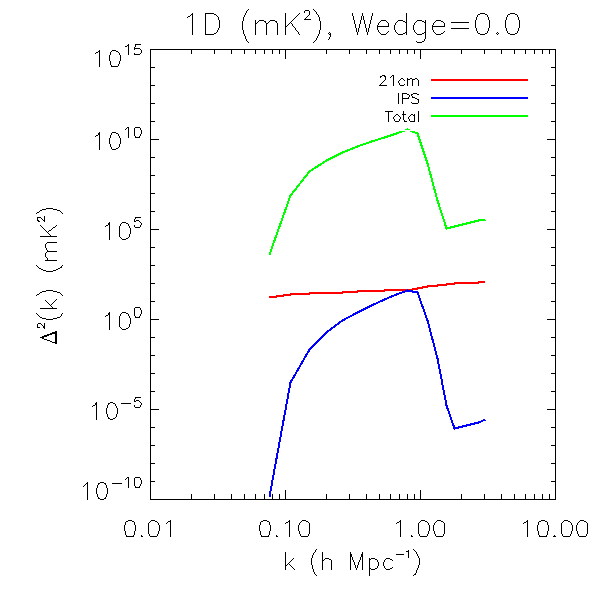}
}\\
\subfigure{
\includegraphics[width=.36\textwidth]{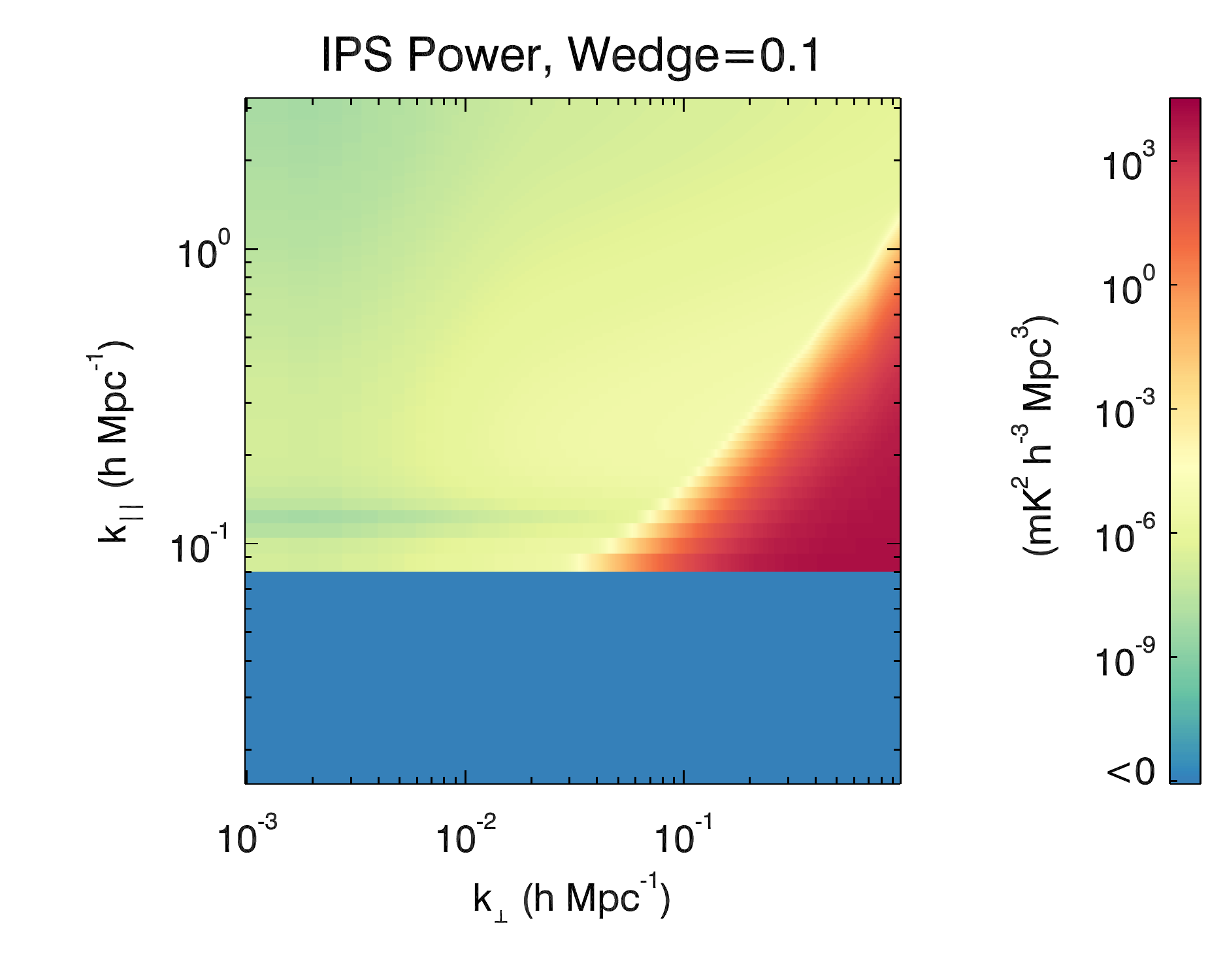}
}
\subfigure{
\includegraphics[width=.28\textwidth]{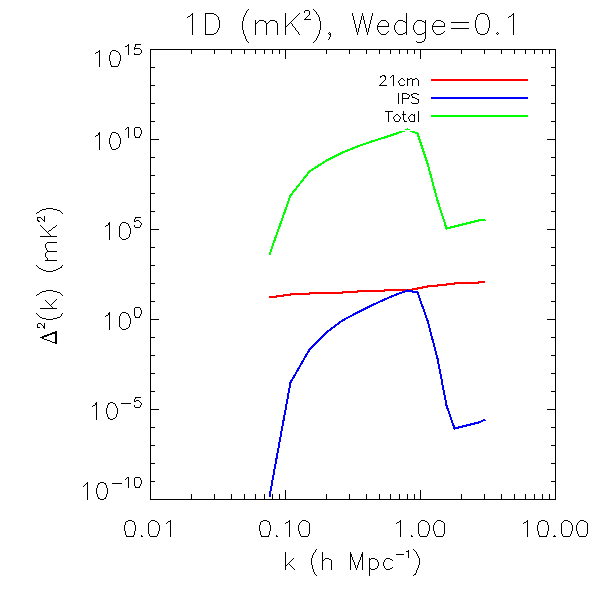}
}\\
\subfigure{
\includegraphics[width=.36\textwidth]{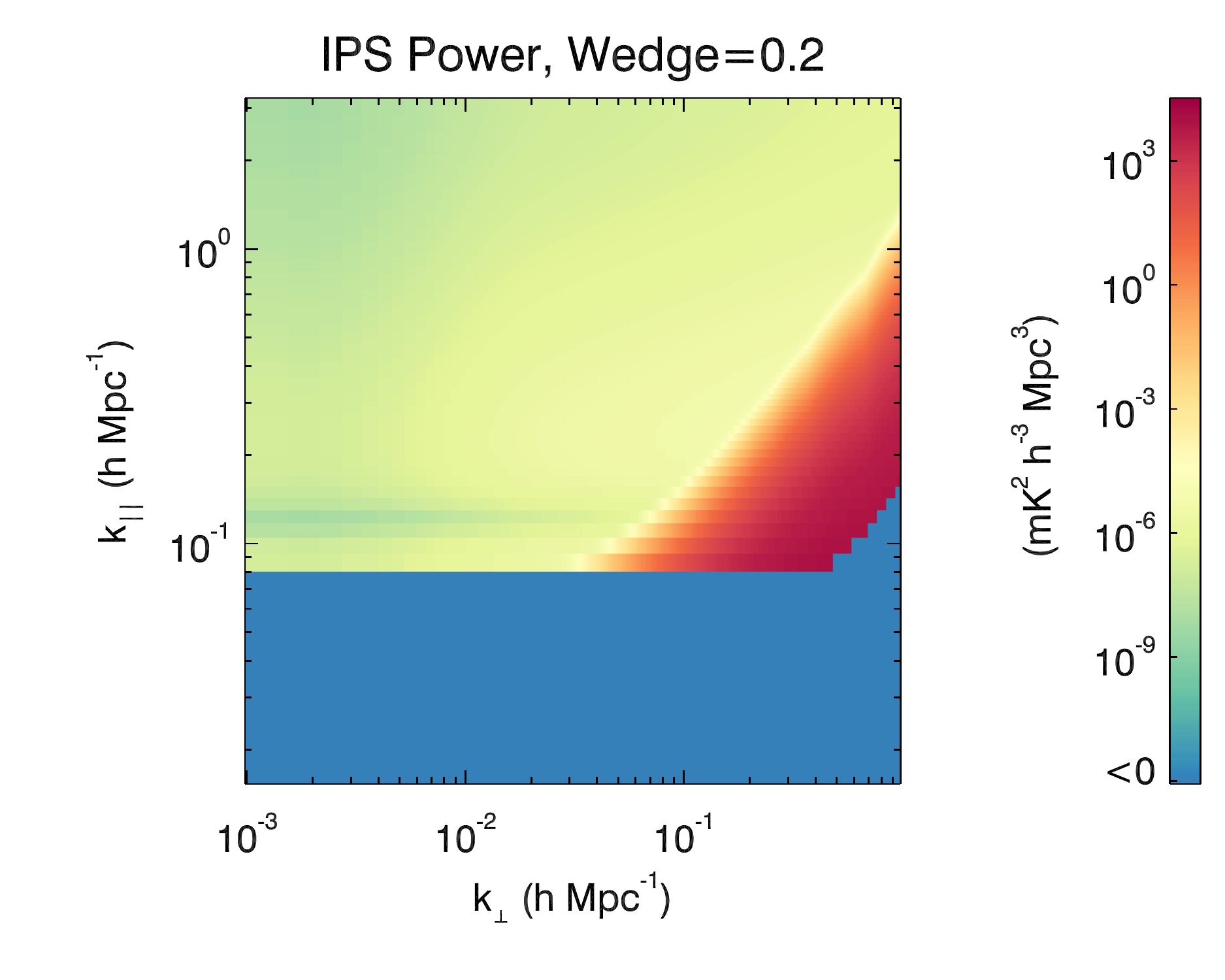}
}
\subfigure{
\includegraphics[width=.28\textwidth]{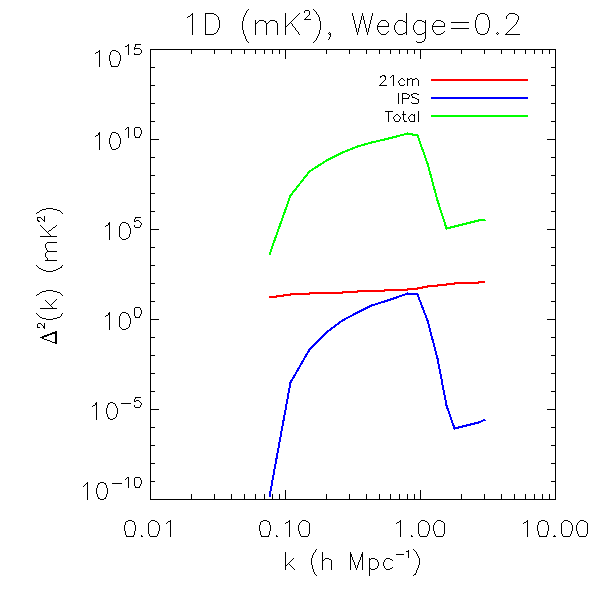}
}\\
\subfigure{
\includegraphics[width=.36\textwidth]{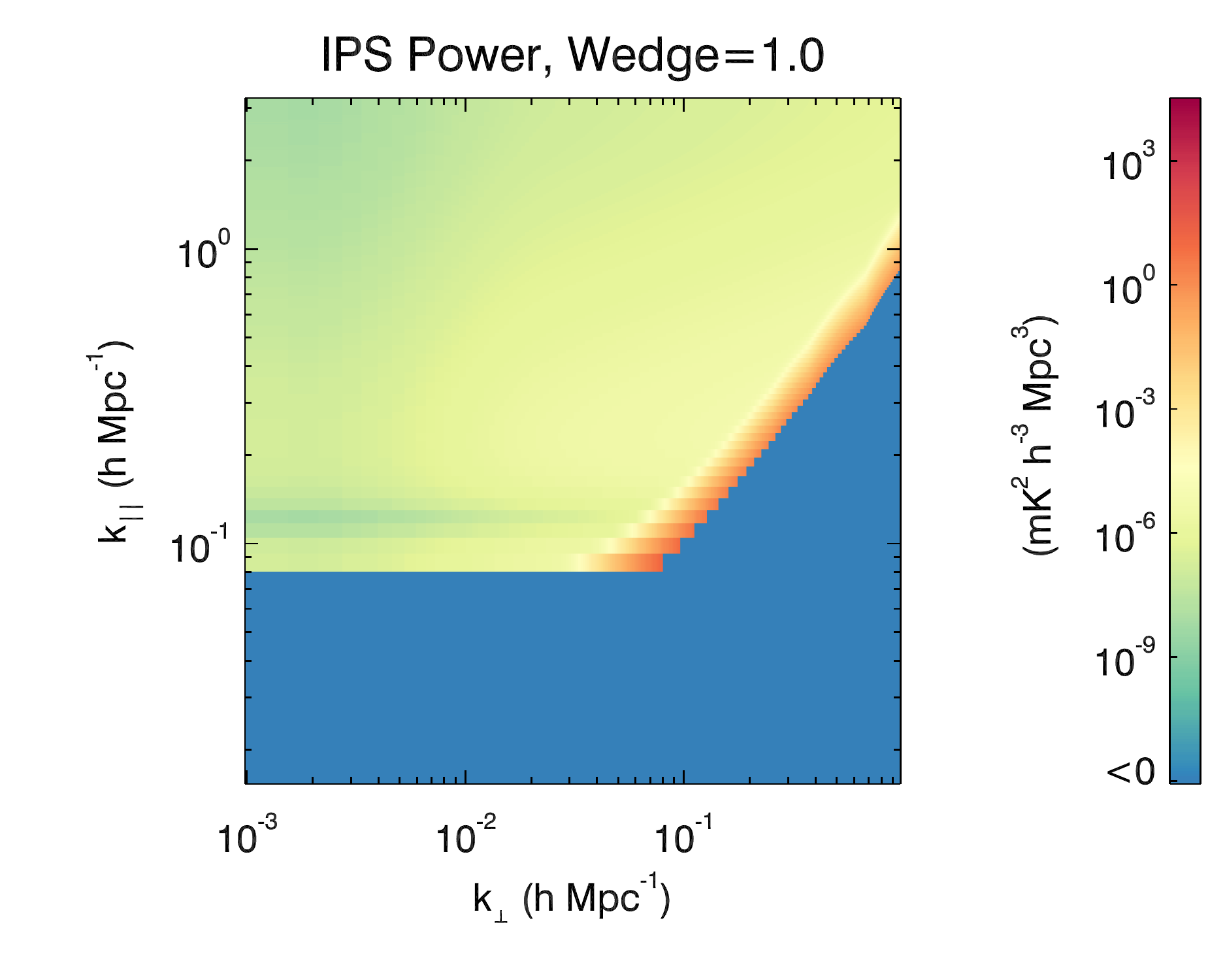}
}
\subfigure{
\includegraphics[width=.28\textwidth]{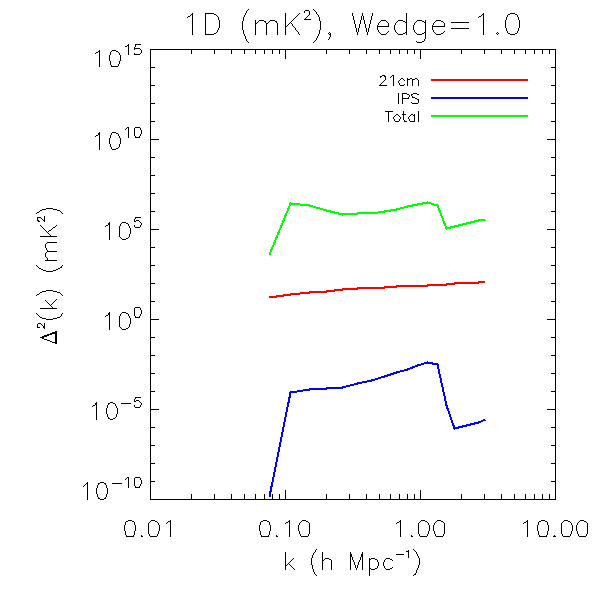}
}
\caption{(Left) 2D IPS power spectra with a region excised (blue mask). (Right) Corresponding spherically-averaged dimensionless power spectra when some power is excised, including the `Total' foreground power, where $P_{\rm tot} = P_{\rm FG} + P_{\rm IPS}$.}
\label{fig:avoidance}
\end{figure}

We use Equation \ref{equation:1d} to average the cosmological power and IPS power to 1D, and compare the amplitudes as a function of spatial wavenumber. To address the impact for different foreground treatment options (e.g., avoidance and suppression), we perform this for the full 2D dataset, as well as for the data with areas of parameter space excised. Figure \ref{fig:avoidance} shows four different excision regions (including no excision), where the `wedge' (W) excises the region with $k_\bot > Wk_\parallel$. (Note that here the sharp drop in IPS power is due to the edge of the sampled region in $k_\bot$ artifically cutting the power. Although this depends on the extent of the simulation performed here, higher $k_\bot$ modes are not sampled in a real EoR experiment, where smaller $k$ modes are used.) Also plotted is the total foreground power ($P_{\rm tot} = P_{\rm FG} + P_{\rm IPS}$), for reference. This reference is particularly useful to compare, because both the IPS power and static foreground power have had the same bandpass taper function applied.
\pagebreak

\begin{figure}[t]
\centering
\subfigure{
\includegraphics[width=.36\textwidth]{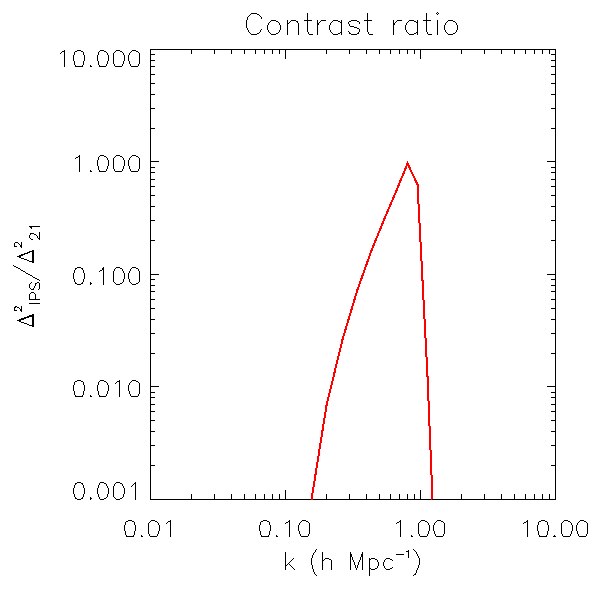}
}
\subfigure{
\includegraphics[width=.36\textwidth]{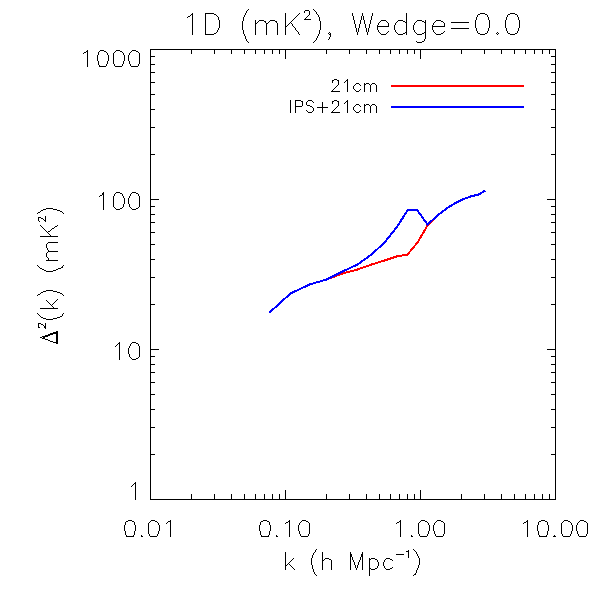}
}
\caption{(Left) Contrast between power in IPS and a model cosmological signal when no power is excised. (Right) Corresponding bias when IPS power is added to cosmological power.}
\label{fig:wedgezero}
\end{figure}

When no power is excised, the IPS power reaches the expected cosmological signal. This is demonstrated more clearly in Figure \ref{fig:wedgezero} where the constrast of IPS contamination to signal (left) and the bias in the measured cosmological power spectrum when IPS adds power (right) is shown. 
However, when only a small component of the foreground wedge region is excised ($k_\bot > 0.2k_\parallel$), the IPS power falls five orders of magnitude below the signal. This indicates that foreground avoidance approaches will be successful in avoiding this effect. For foreground suppression approaches, the additional power \textit{should} be included in the foreground data model. This is particularly true when one considers the different spectral structure of IPS compared with static extragalactic foregrounds. This conclusion will again be dependent on the bandpass taper function, with the exact location of the `edge' of the wedge, and the leakage changing.

\subsubsection*{Anisotropy}
\begin{figure}
\plotone{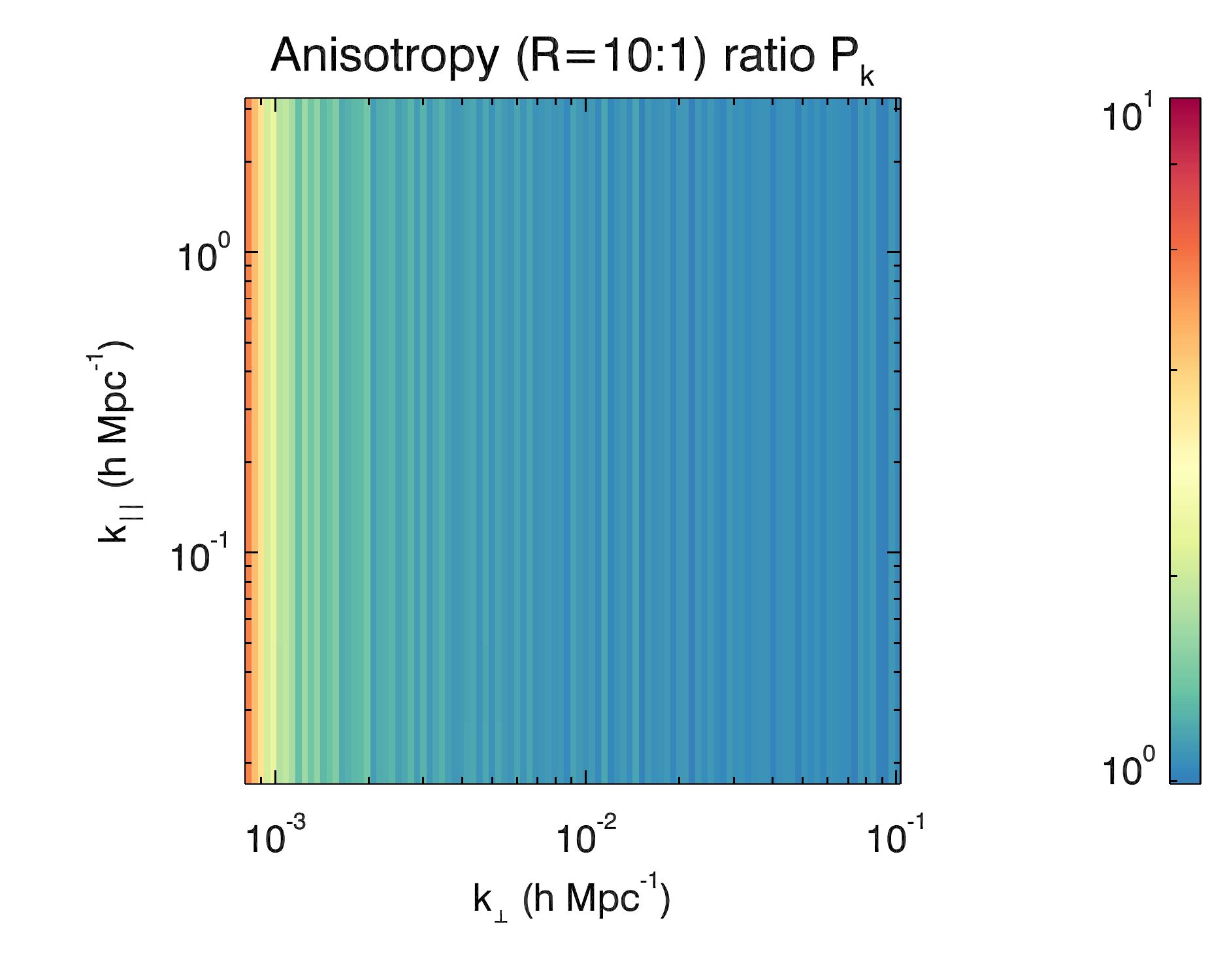}
\caption{Ratio of the IPS power for an input anisotropic power spectrum with $R=10$, to that for an isotropic power spectrum.\label{fig:anisotropy}}
\end{figure}

We computed the impact of IPS for different elongations of the solar wind turbulence. An anisotropic electron density power spectrum was found to have a minimal impact on the IPS power spectrum. We considered axial ratios of $R={1.0,2.0,5.0,10.0}$ to bracket the range empirically deduced from IPS and ISS observations \citep{dennison69,macquart07}. Ratios of IPS power in the $k-k$ parameter space demonstrated deviations from unity that were contained in very low angular $k$ modes (large angular scales). Again, the dampening effects of the Fresnel filter and temporal sampling washed-out any significant deviations at smaller scales. Figure \ref{fig:anisotropy} shows the ratio of IPS power for an input power spectrum with $R=10$ to the isotropic case considered above.
Even for this large ratio the insignificant impact across the parameter space does not warrant any additional treatment of anisotropic power spectra compared with isotropy. Importantly, additional spectral structure is not introduced. The apparent variations in $k_\bot$ are not real, and correspond to differences in the binning.

\section{Discussion and conclusions}\label{sec:conc}
The results of Figure \ref{fig:signalpower} predict that IPS-induced intensity fluctuations will not impact detectability of the EoR signal if one assumes a foreground avoidance strategy. However, if an EoR estimator works within the wedge, the additional power due to IPS above that expected for static foregrounds at their mean flux density, and the cosmological signal, will need to be taken into account. In this case, one may try to measure the flux densities of fainter sources than would otherwise be undertaken, in order to identify IPS effects. One may also use the results from brighter sources, assuming they are distributed across the FOV, to identify any IPS-related intensity fluctuations and treat these in the analysis. For foreground suppression approaches, one may include an additional term in the foreground data covariance matrix that applies the IPS statistical model developed here to the estimator \citep[e.g., like that used in ][]{dillon15}. The key component here beyond an increase in the amplitude of the extragalactic foreground power, is the different spatial and spectral structure of IPS.

The power spectrum of intensity fluctuations is heavily affected by the Fresnel filter. At EoR observation frequencies, the Fresnel frequency is $\nu_F(150{\rm MHz})\simeq{0.36{\rm Hz}}$, corresponding to suppression of the power below $k_\perp\simeq{10^2}{\rm Mpc}^{-1}$. Therefore, the large angular scales of relevance to the EoR experiments, are contained entirely within the region of Fresnel filter suppression. This has the effect of dampening the sharp large-scale plasma density fluctuations spectrum ($n\simeq{3.5}$), resulting in significantly less power than would otherwise occur. The sinc suppression of small-scale power (high $k_x$) due to the temporal averaging from the observation time, affects modes with $k_x\gtrsim{30}{\rm Mpc}^{-1}$. The two effects therefore combine to suppress power across all spatial scales. Inclusion of anisotropy in the power spectrum did not have any measurable impact. It is worth noting that interstellar scintillation (ISS) will also induce intensity variation in extragalactic point sources, but there the large distance to the phase screen implies a substantially longer decorrelation timescale ($\sim$months).

These results have assumed the weak scattering regime at large solar elongations. This will typically be the case for EoR observations, which are not performed close to the Sun. They also assume a fixed distance to the plasma (1.AU) and scintillation index, $m=1$. The IPS power can be scaled directly with these parameters, such that:
\begin{equation}
P_{\rm obs}(L,m) = P_{\rm obs} \left( \frac{m^2}{1.0} \right) \left( {\frac{1.{\rm AU}}{L}}\right)^2,
\end{equation}
where the distance scaling converts linear to angular scales. With these scalings, it is unlikely that an unusually large intensity fluctuation, or unusually nearby plasma, will affect the EoR estimation, except if working within the contaminated wedge region.

\bibliographystyle{jphysicsB}
\bibliography{pubs.bib}

\acknowledgements
We would like to thank Jean-Pierre Macquart and Ron Ekers for fruitful discussions, and the referee for helping to make the manuscript accessible to EoR and IPS communities.
This research was supported under Australian Research Council's Discovery Early Career Researcher funding scheme (project number DE140100316) and the Centre for All-sky Astrophysics (an Australian Research Council Centre of Excellence funded by grant CE110001020). We acknowledge the iVEC Petabyte Data Store, the Initiative in Innovative Computing and the CUDA Center for Excellence sponsored by NVIDIA at Harvard University, and the International Centre for Radio Astronomy Research (ICRAR), a Joint Venture of Curtin University and The University of Western Australia, funded by the Western Australian State government.

\end{document}